\documentclass[aps,prb,twocolumn,superscriptaddress,showpacs,floatfix]{revtex4}
\usepackage{graphicx}
\usepackage{dcolumn}
\usepackage{bm}
\usepackage{stmaryrd}
\usepackage{latexsym}
\usepackage{amssymb}
\usepackage{amsfonts}
\usepackage{amsmath}

\begin{document}

\title{Interacting Fermi Gases in Disordered One-Dimensional Lattices}
\author{Gao Xianlong}
\email{x.gao@sns.it}
\affiliation{NEST-CNR-INFM and Scuola Normale Superiore, I-56126 Pisa, Italy}
\author{M. Polini}
\email{m.polini@sns.it}
\affiliation{NEST-CNR-INFM and Scuola Normale Superiore, I-56126 Pisa, Italy}
\author{B. Tanatar}
\affiliation{Department of Physics, Bilkent University, Bilkent, 06800 Ankara, Turkey}
\author{M.P. Tosi}
\affiliation{NEST-CNR-INFM and Scuola Normale Superiore, I-56126 Pisa, Italy}

\date{\today}
\begin{abstract}
Interacting two-component Fermi gases loaded in a one-dimensional ($1D$) 
lattice and subject to harmonic trapping exhibit intriguing compound phases in 
which fluid regions coexist with local Mott-insulator and/or 
band-insulator regions.
Motivated by experiments on cold atoms inside 
disordered optical lattices, we present a theoretical study of the 
effects of a random potential on these ground-state phases. Within a density-functional 
scheme we show that disorder has two main effects: (i) 
it destroys the local insulating regions if it is 
sufficiently strong compared with the on-site atom-atom repulsion, and 
(ii) it induces an anomaly in the compressibility at low density from quenching of percolation.
\end{abstract}
\pacs{03.75.Lm, 03.75.Ss, 71.30.+h, 71.10.Pm}
\maketitle

{\it Introduction} ---The interplay between interactions and disorder 
in quantum many-body systems is an area of long-standing 
interest. For instance, both long-ranged Coulomb interactions and 
disorder from various mechanisms are believed to play an important 
role in the metal-insulator transition (MIT) in the 
two-dimensional ($2D$) electron liquid in semiconductor heterostructures~\cite{MIT}. 
Disorder and interactions affect not only transport properties of the $2D$ liquid, but also thermodynamic quantities such as the compressibility~\cite{compressibility_anomaly,droplet_state} and the 
spin susceptibility~\cite{spin_susceptibility}. ``Dirty-boson" systems such as 
liquid $^{4}{\rm He}$ absorbed in aerogel, Vycor, or Geltech~\cite{helium_four}, or disordered granular superconductors~\cite{fazio_2001}, have also been extensively studied.

Cold atom gases are becoming important tools to 
understand the interplay between single-particle randomness and 
cooperative effects such as superfluidity 
and many-body effects induced by interactions~\cite{speckle}. 
Atoms trapped in an optical lattice (OL) are particularly suitable candidates for such studies, especially 
because they allow one to reach the strongly-coupling regime 
through the depression of the kinetic energy associated with 
well-to-well tunneling~\cite{optical_lattices}. 
A $^{87}{\rm Rb}$ Bose-Einstein condensate (BEC) inside a disordered $1D$ OL has been used to study the interplay between repulsive interactions and disorder~\cite{hannover}.
In this work it has also been pointed out that thermodynamic quantities, such as the superfluid density in the case of a BEC, provide a better indicator of disorder-induced localization than time-of-flight absorption images. 
The present work has been motivated by the experiments in Refs.~\onlinecite{speckle} and~\onlinecite{hannover}. 
We report a study of the interplay between interactions 
and randomness in a repulsive two-component Fermi gas trapped in a $1D$ OL. 
An added motive of interest is that the Landau Fermi-liquid paradigm does not 
apply~\cite{luttinger_liquids}. Two-component Fermi gases 
have recently been prepared in a quasi-$1D$ 
geometry~\cite{moritz_2005}, thus opening the way to experimental studies of $1D$ phenomena such 
as spin-charge separation~\cite{recati_2003} and 
atomic-density waves~\cite{gao_ADWs}. 

\begin{figure*}
\begin{center}
\includegraphics[width=0.80\linewidth]{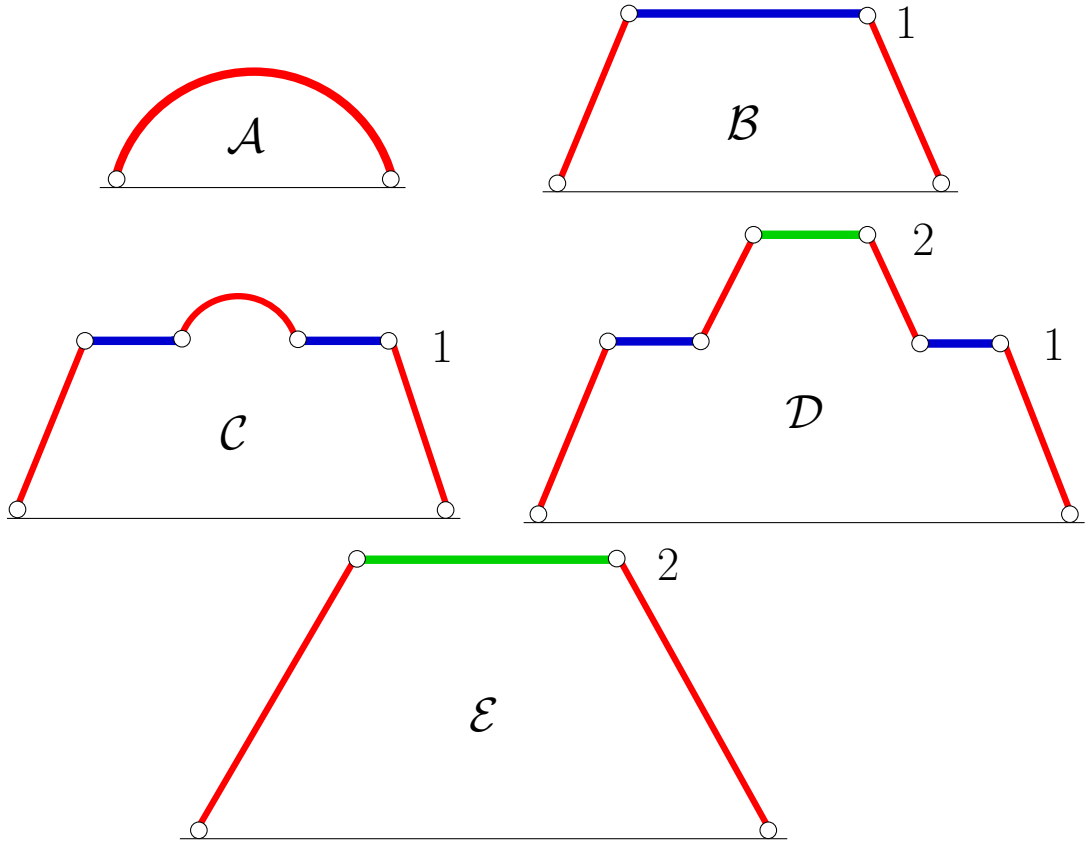}
\caption{(color online) Sketch of the ground-state site occupation $n_i$ of an 
interacting Fermi gas in a harmonic trap and a 
clean $1D$ lattice. Phase ${\cal A}$ is a fluid (known as the metallic phase in the case of electrons) 
with $0<n_i<2$. In phase 
${\cal B}$ an incompressible Mott insulator occupies the bulk of the trap 
with $n_i$ locally locked to $1$. In phase ${\cal C}$ a fluid 
with $1<n_i<2$ is embedded in the Mott plateau. In phase 
${\cal D}$ a band insulator with $n_i$ locally 
locked to $2$ is surrounded by fluid edges and embedded in the 
Mott plateau. Finally, in phase ${\cal E}$ a band insulator in 
the bulk of the trap coexists with fluid edges.\label{fig:one}}
\end{center}
\end{figure*}

The ground state of an interacting Fermi gas moving under 
harmonic confinement in a $1D$ OL shows in the 
absence of disorder five qualitatively different 
phases~\cite{rigol_2003,drummond_2005,capelle_2005,gao_long_2005} 
(for a pictorial description see Fig.~\ref{fig:one}). 
How does disorder influence these phases and their thermodynamic properties? 
In the following we provide a quantitative answer to this question. 
In particular, we demonstrate that the incompressible 
Mott-insulating regions are very stable against 
disorder at strong coupling. 
We also show that the compressibility exhibits a disorder-induced 
low-density anomaly, similar in some 
respects to the one which has been found 
both experimentally~\cite{compressibility_anomaly} and theoretically~\cite{droplet_state} in the $2D$ electron liquid
close to the onset of the MIT.

{\it The $1D$ random Fermi-Hubbard model} ---
We consider a two-component Fermi gas 
with $N_{\rm f}$ atoms constrained to move 
under harmonic confinement of strength $V_2$ inside a 
disordered $1D$ OL with unit lattice constant and $N_{\rm s}$ 
lattice sites $i\in[1,N_{\rm s}]$. The system is 
described by a single-band Hubbard Hamiltonian,
\begin{eqnarray}\label{eq:hubbard}
{\hat {\cal H}}&=&-\sum_{i,j}\sum_\sigma
t_{ij}({\hat c}^{\dagger}_{i\sigma}{\hat c}_{j\sigma}+{\rm H}.{\rm c}.)+
U\sum_{i}\,{\hat n}_{i\uparrow}{\hat n}_{i\downarrow}\nonumber\\
&+&V_2\sum_{i}(i-N_{\rm s}/2)^2{\hat n}_i+\sum_{i}\varepsilon_i\,{\hat n}_i\,.
\end{eqnarray}
Here $t_{ij}=t>0$ if $i,j$ are nearest sites and zero 
otherwise, $\sigma=\uparrow,\downarrow$ is 
a pseudospin-$1/2$ label for two internal 
hyperfine states, ${\hat n}_{i\sigma}= {\hat c}^{\dagger}_{i\sigma}{\hat c}_{i\sigma}$ 
is the pseudospin-resolved site occupation operator, and 
${\hat n}_i=\sum_\sigma {\hat n}_{i\sigma}$. The effect of disorder is 
simulated by the last term in Eq.~(\ref{eq:hubbard}), where 
${\varepsilon}_i$ is randomly chosen at each site 
from a uniform distribution in the range $[-W/2,W/2]$~~\cite{footnote_1}.
 
In the unconfined limit ($V_2=0$) the Hamiltonian ${\hat {\cal H}}$ reduces to the Anderson localization 
problem~\cite{anderson_1958} for $U=0$ and to the exactly-solvable 
Lieb-Wu model~\cite{lieb_wu} for $W=0$. The Lieb-Wu model describes a 
Luttinger liquid away from half and full filling~\cite{schulz_1990}, 
a Mott insulator at half filling, and a band insulator 
at full filling. In the unconfined limit the $2D$ version of 
${\hat {\cal H}}$ has been studied in connection with the 
$2D$ MIT~\cite{2D_random_Hubbard_model}. 
In the clean limit the ground-state of ${\hat {\cal H}}$ has been studied by several 
authors~\cite{rigol_2003,drummond_2005,capelle_2005,gao_long_2005}, finding five different phases as already shown in Fig.~\ref{fig:one}. 

A particular set $\varepsilon_i(\alpha)$ of randomly chosen values 
of $\varepsilon_i$, labeled by $\alpha$, is a realization of the disorder potential. 
Each realization defines an external potential 
${\cal V}_i(\alpha)\equiv V_2(i-N_{\rm s}/2)^2+\varepsilon_i(\alpha)$ 
which determines, {\it via} site-occupation functional 
theory~\cite{soft}, a site occupation 
$n_i(\alpha)=\langle \Psi_\alpha|{\hat n}_i|\Psi_\alpha\rangle$, 
$|\Psi_\alpha\rangle$ being the ground state (GS) of ${\hat {\cal H}}$ for disorder 
realization $\alpha$. The total energy is a unique functional of 
$n_i(\alpha)$, which can be written as~\cite{soft,gao_long_2005}
\begin{equation}\label{eq:totenergy}
{\cal E}[n(\alpha)]={\cal F}_{\rm HK}[n(\alpha)]+\sum_{i}{\cal V}_i(\alpha)n_i(\alpha)\,.
\end{equation}
Here ${\cal F}_{\rm HK}[n(\alpha)]$ is a {\it universal} functional 
of the site occupation, in the sense that it does not depend on the 
external potential ${\cal V}_i(\alpha)$. The GS site occupation 
can be found by solving the Euler-Lagrange equation
\begin{equation}\label{eq:HK_variational}
\frac{\delta {\cal F}_{\rm HK}[n(\alpha)]}{\delta n_i(\alpha)}+{\cal V}_i(\alpha)=\mu(\alpha)\,,
\end{equation}
the constant $\mu(\alpha)$ being a Lagrange multiplier to enforce 
particle conservation, {\it i.e.} $\sum_i n_i(\alpha)=N_{\rm f}$. 
A local-density approximation (LDA) will
be used below for ${\cal F}_{\rm HK}[n(\alpha)]$.

The site occupation ${\cal N}_i$ due to 
${\hat {\cal H}}$ is finally obtained by means of a disorder ensemble average, 
{\it i.e.} ${\cal N}_i=\langle\langle n_i\rangle\rangle_{\rm dis}$ where
$
\langle\langle {\cal O}\rangle\rangle_{\rm dis}=
\lim_{{\cal M}\rightarrow \infty}\frac{1}{\cal M}
\sum_{\alpha=1}^{\cal M}{\cal O}(\alpha)
$.
In practice one can average only over a finite number ${\cal M}$ 
of disorder realizations. Due to our very efficient computational  
method~\cite{gao_long_2005} we are easily able to average over 
${\cal M}=10^4$ realizations of disorder for every set of 
parameters $\{N_{\rm s},N_{\rm f},u=U/t,V_2/t,W/t\}$. 
Averaging over such large numbers of realizations becomes
necessary as the strength of disorder increases and we have checked that the density profiles that we report below are stable against further increases of ${\cal M}$. Finally, the global compressibility 
can be obtained from the stiffness ${\cal S}_\rho=\langle\langle
\delta \mu/\delta N_{\rm f}\rangle\rangle_{\rm dis}$ (see Ref.~\onlinecite{luttinger_liquids}).

{\it Site occupation and stiffness anomaly in the presence of disorder} ---
The functional ${\cal F}_{\rm HK}$ is the sum of 
three terms,
\begin{equation} 
{\cal F}_{\rm HK}[n(\alpha)]=T_s[n(\alpha)]+\frac{U}{2}\sum_i n^2_i(\alpha)
+{\cal E}_{\rm xc}[n(\alpha)]\,.
\end{equation} 
The first term is the noninteracting kinetic energy functional, 
which is approximated in this work 
``\`a la Thomas-Fermi"~\cite{capelle_2005,gao_long_2005}, 
{\it i.e.} through an LDA based on 
the noninteracting kinetic energy of the Lieb-Wu model. The other terms 
in ${\cal F}_{\rm HK}$ are the mean-field interaction energy and 
the exchange-correlation energy functional incorporating many-body 
effects beyond mean field. This is approximated 
through an LDA based on the exactly known exchange-correlation 
energy of the Lieb-Wu model~\cite{gao_long_2005}. 
Within this LDA an explicit equation for $n_i(\alpha)$ can easily be 
derived~\cite{capelle_2005,gao_long_2005} 
from Eq.~(\ref{eq:HK_variational}). 
Its self-consistent solution leads
to the site occupation $n_i(\alpha)$ and to the chemical potential 
$\mu(\alpha)$ for a particular disorder realization~\cite{footnote_2}, and hence to the disorder-ensemble averaged quantities.

\begin{figure*}
\begin{center}
\tabcolsep=0cm
\begin{tabular}{c}
\includegraphics[width=0.80\linewidth]{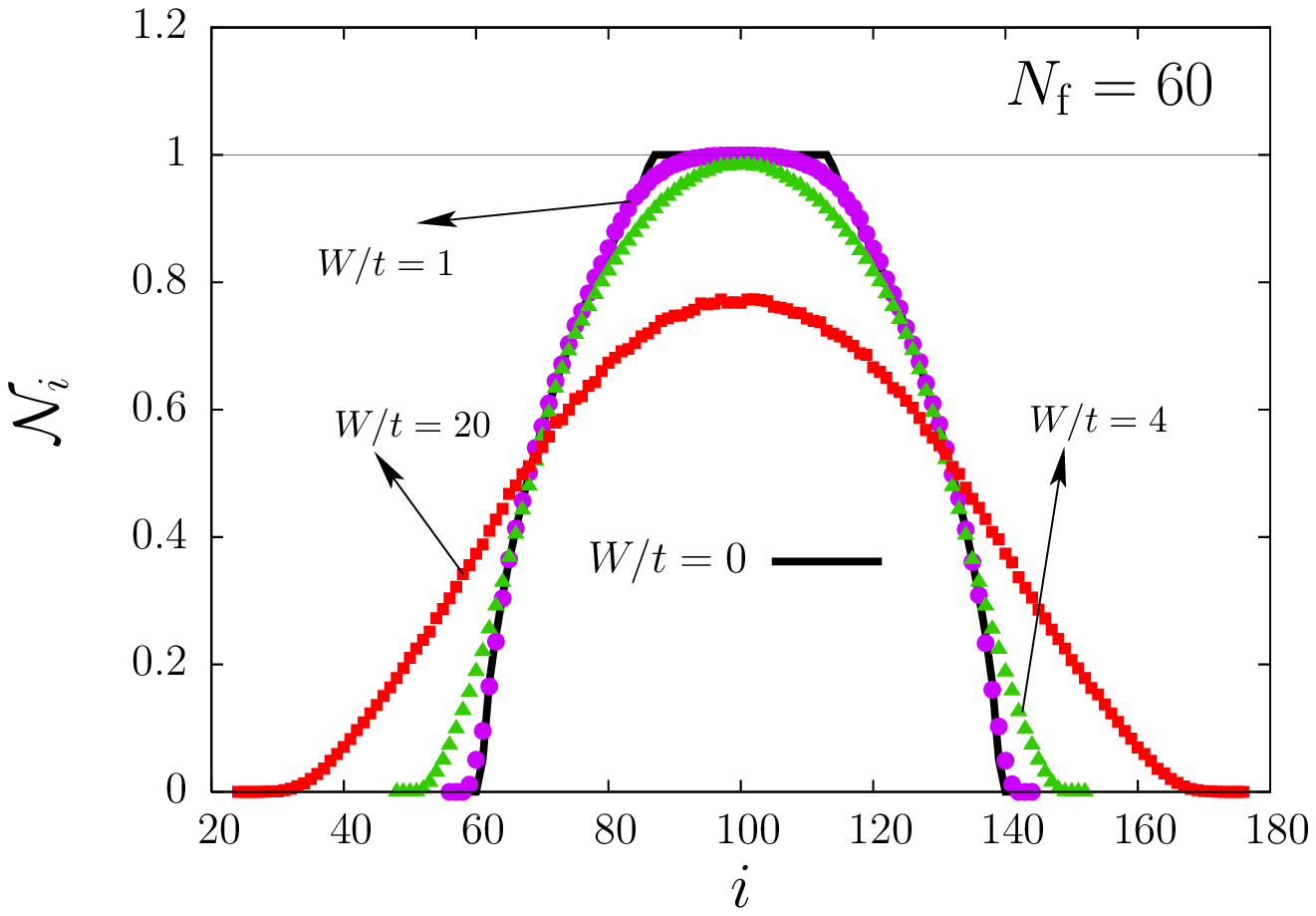}\\
\includegraphics[width=0.80\linewidth]{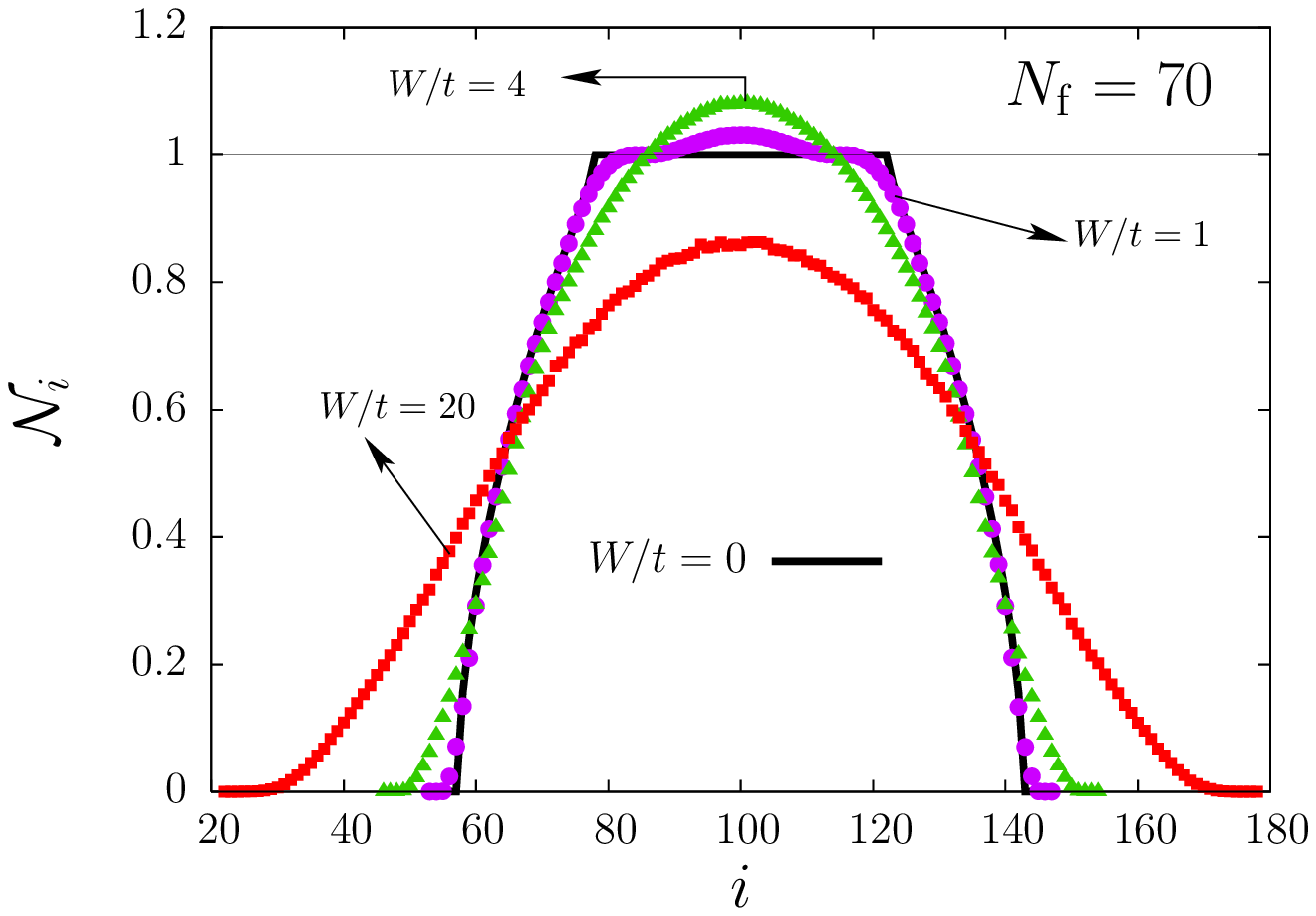}
\end{tabular}
\caption{(color online) Site occupation ${\cal N}_i$ as a function 
of site position $i$ for $u=4$ and $V_2/t=2.5\times 10^{-3}$ in a lattice with 
$N_{\rm s}=200$ sites. The number of atoms is 
$N_{\rm f}=60$ in the top panel and $N_{\rm f}=70$ in the 
bottom panel.\label{fig:two}}
\end{center}
\end{figure*}

In Fig.~\ref{fig:two} we show the site occupation ${\cal N}_i$ for a disordered 
gas with $u=4$, which is in phase ${\cal B}$ for $W=0$. 
The edges of the Mott plateau are the first to be corrupted 
by the appearence of weak disorder. The stability of the central region of the plateau 
depends on the number of atoms, if all other parameters are 
kept fixed. For $N_{\rm f}=60$ (top panel of Fig.~\ref{fig:two}) the central region 
persists over a finite range of disorder. Disorder increases 
the tunneling through the edges of the trap, making the confinement 
effectively weaker, and thus leads to broadening of the 
site occupation. The Mott plateau has disappeared at 
$W/t=4$, and for $W/t=20$ the site-occupation profile has a strongly non-parabolic overall shape. 
For $N_{\rm f}=70$ instead (bottom panel of Fig.~\ref{fig:two}), the Mott plateau at the 
center of the trap is unstable against the formation of a 
fluid phase with ${\cal N}_i>1$, and the Mott phase can survive for weak disorder only in an intermediate region between the edge and the center of the trap. In fact, $N_{\rm f}=70$ is the critical number of atoms at which the phase transition ${\cal B}\rightarrow {\cal C}$ occurs in the clean limit: a weak disorder potential can shift the transition and induce a fluid region embedded in the Mott plateau (see the plot corresponding to $W/t=1$ in the bottom panel of Fig.~\ref{fig:two}). Eventually, the scenario depicted above for 
$N_{\rm f}=60$ is re-established when $W$ is strong enough.

\begin{figure*}
\begin{center}
\tabcolsep=0cm
\begin{tabular}{c}
\includegraphics[width=0.80\linewidth]{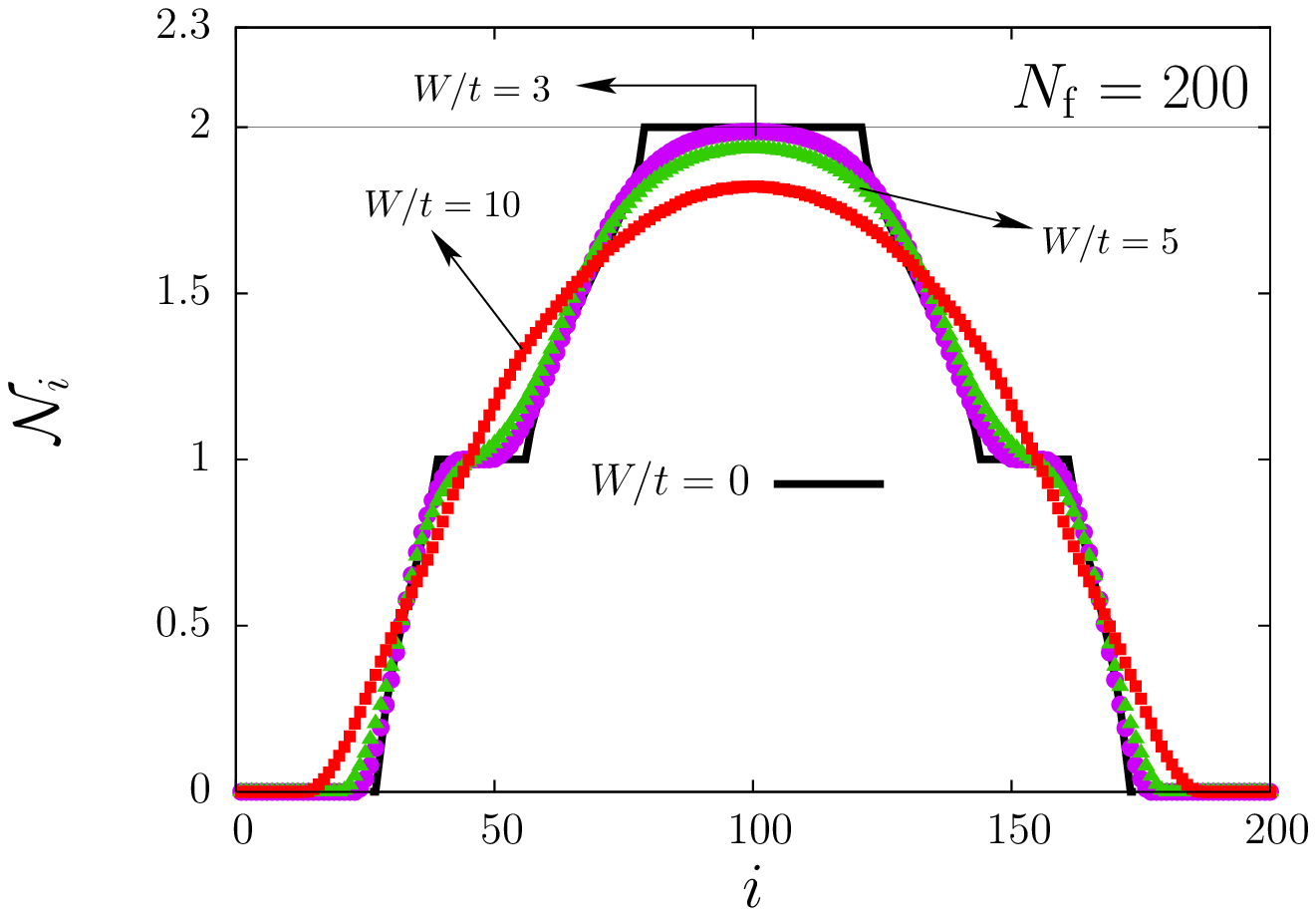}\\
\includegraphics[width=0.80\linewidth]{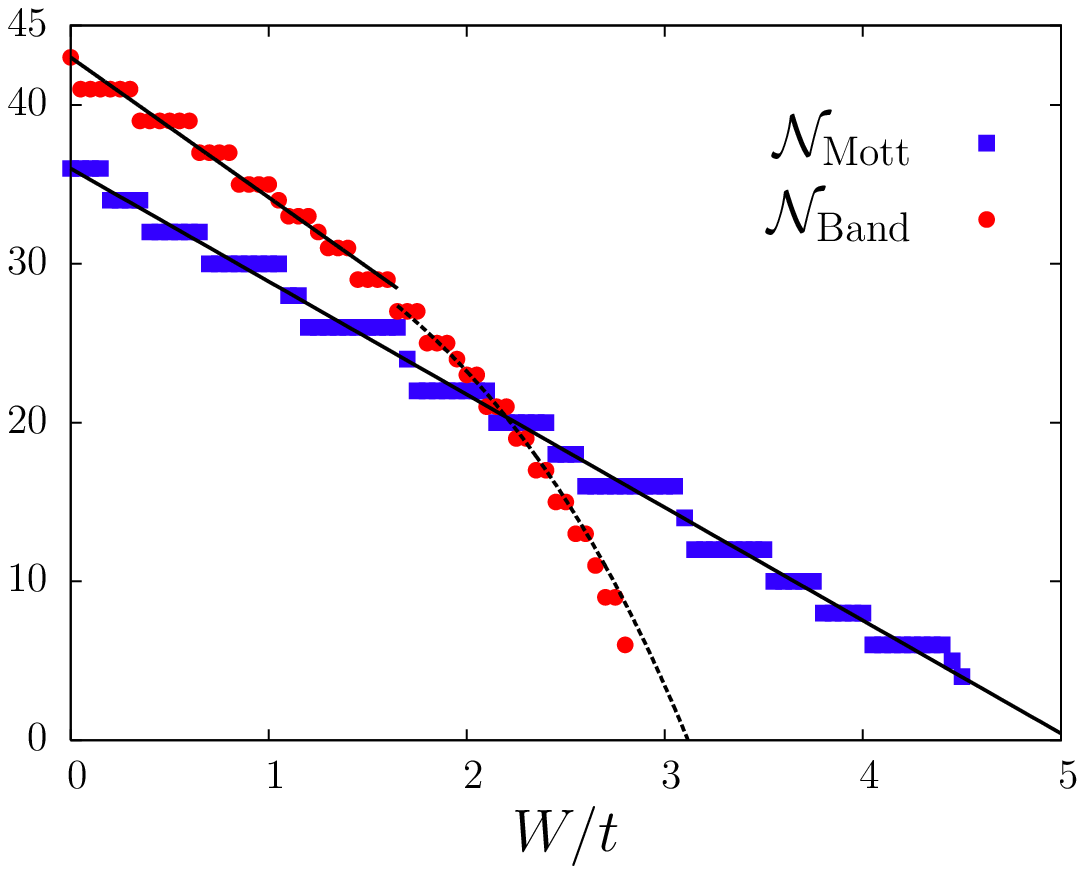}
\end{tabular}
\caption{(color online) Top panel: Site occupation ${\cal N}_i$ 
as a function of $i$ for $N_{\rm f}=200$ atoms in a lattice with $N_{\rm s}=200$ sites, 
in the case $u=8$ and $V_2/t=2.5\times 10^{-3}$. 
Bottom panel: Number of consecutive sites ${\cal N}_{\rm Mott}$ 
(${\cal N}_{\rm Band}$) 
at which $|{\cal N}_i-1|\leq 10^{-5}$ ($|{\cal N}_i-2|\leq 10^{-5}$), 
as a function of $W/t$ for the system shown in the top panel. 
The steps show that each insulating region 
is stable over a finite range of $W/t$. 
${\cal N}_{\rm Mott}$ can be fitted with a linear function over the whole range of $W/t$ 
(solid line through the squares, showing that ${\cal N}_{\rm Mott}=0$ at $(W/t)_{{\rm c}1}\approx 5$). 
${\cal N}_{\rm Band}$ has a linear behavior up to
$W/t \approx 1.8$ and beyond can only be fitted by a nonlinear 
function (solid and dashed lines through the dots, showing that ${\cal N}_{\rm band}=0$ at 
$(W/t)_{{\rm c}2}\approx 3 <(W/t)_{{\rm c}1}$).\label{fig:three}}
\end{center}
\end{figure*}

In the top panel of Fig.~\ref{fig:three} we show the site occupation 
for a disordered gas with $u=8$, which is in phase ${\cal D}$ 
for $W=0$. Clearly, the band-insulating 
region can only be corrupted from below. We note that at 
$W/t=3$ the Mott-insulating regions still exist, while the 
band-insulating region has been destroyed. 
This confirms the expectation that a Mott-insulating region, 
having its origin in exchange-correlation effects, 
is more stable against disorder than a band-insulating region. 
These behaviors are examined in detail in the bottom panel of Fig.~\ref{fig:three}.

\begin{figure*}
\begin{center}
\tabcolsep=0cm
\begin{tabular}{c}
\includegraphics[width=0.80\linewidth]{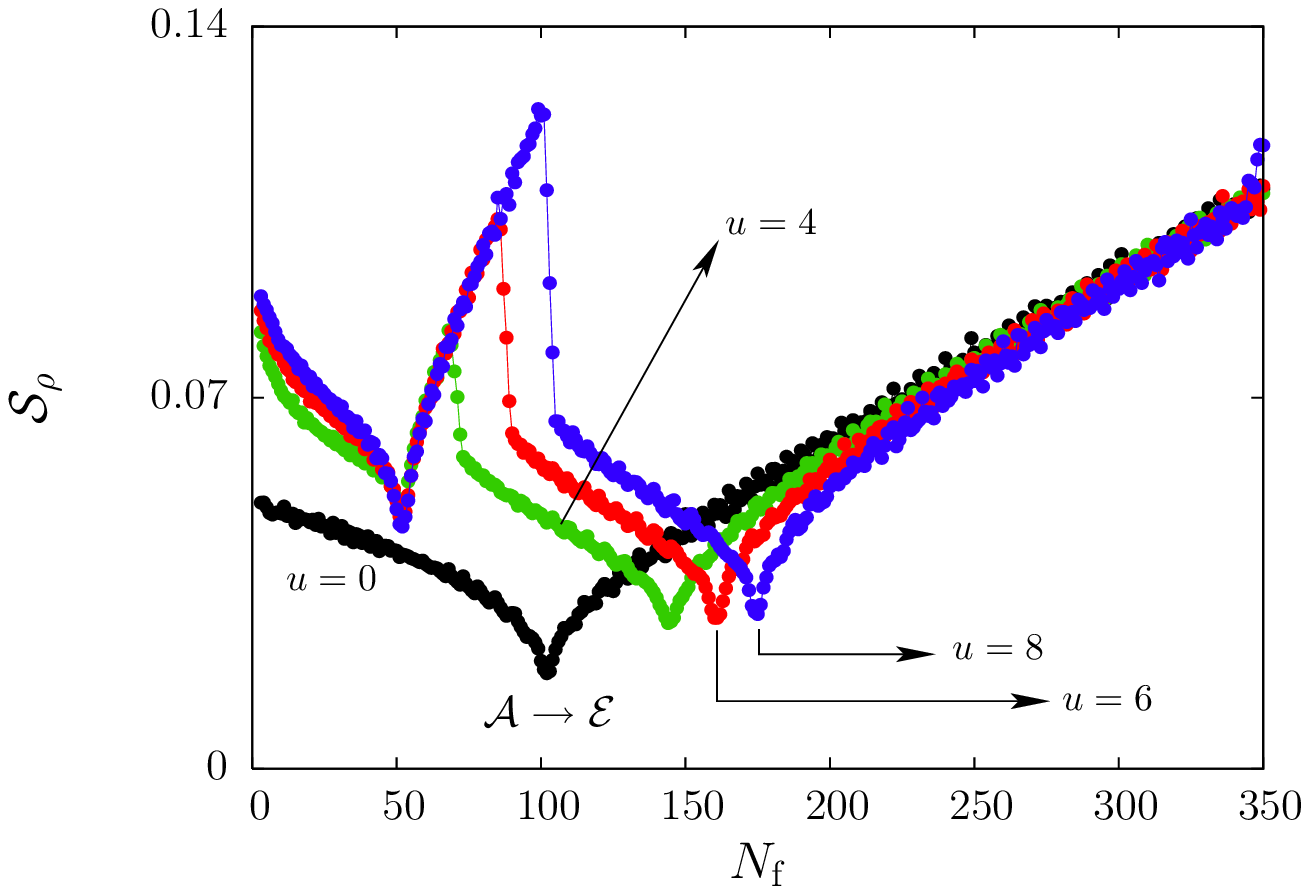}\\
\includegraphics[width=0.80\linewidth]{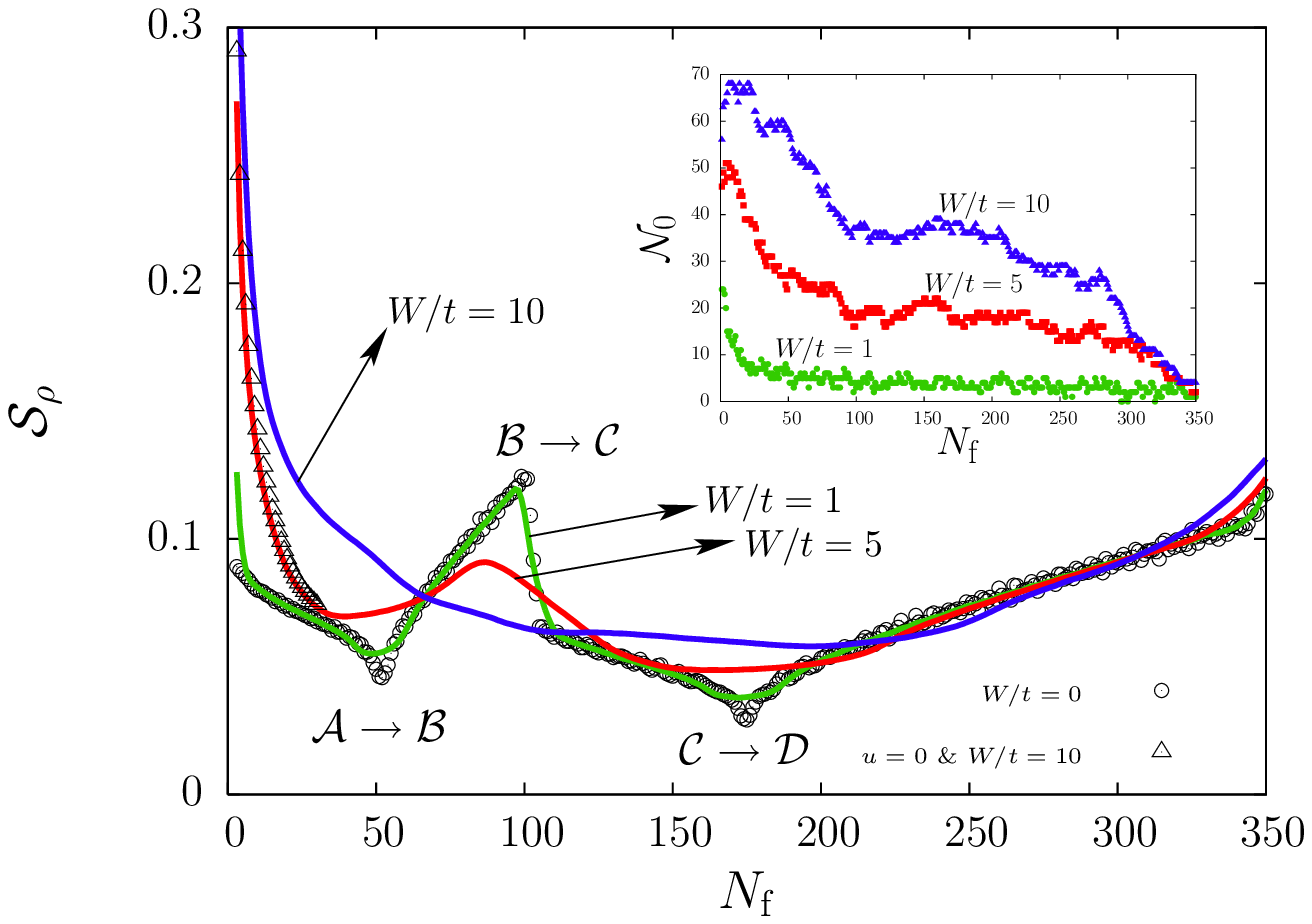}
\end{tabular}
\caption{(color online) Thermodynamic stiffness ${\cal S}_\rho$ (in units of $t$) 
as a function of $N_{\rm f}$ for $V_2/t=2.5\times 10^{-3}$ and $N_{\rm s}=200$ lattice sites.
Top panel: results for a clean system at various values of $u$ (in the noninteracting case only the phase transition ${\cal A}\rightarrow {\cal E}$ can occur). Bottom panel: results for a disordered system at $u=8$ and for $0\leq W/t \leq 10$. The black triangles report the low-density stiffness of a noninteracting system at $W/t=10$. The inset shows the number ${\cal N}_0$ of sites at which $n_i(\alpha)\leq 10^{-5}$ in a particular realization of disorder, as a function of $N_{\rm f}$ for $u=8$ and $0\leq W/t \leq 10$.\label{fig:four}}
\end{center}
\end{figure*}

We turn in Fig.~\ref{fig:four} to illustrate the effect of disorder on the stiffness 
of the Fermi gas. In the top panel we show ${\cal S}_\rho$ as a function of 
$N_{\rm f}$ at different 
values of $u$ in the absence of disorder. At $u\geq 4$ this quantity exhibits three non-analyticity 
points associated with the three 
phase transitions: ${\cal A}\rightarrow {\cal B}$, 
${\cal B}\rightarrow {\cal C}$ and ${\cal C}\rightarrow {\cal D}$. 
In the bottom panel of Fig.~\ref{fig:four} we show the same disorder-averaged quantity 
for a disordered Fermi gas with $u=8$. 
We see that the disorder has two main effects. 
It not only leads to smoothing of the non-analytic behaviors found in the clean limit, 
but also induces a strong stiffening at 
low density. The latter is an ``anomalous" behavior compared to that found 
in the clean limit. In fact, for finite $W$ the stiffening appears 
to grow unbounded at very low density 
(see bottom panel of Fig.~\ref{fig:four}), following the power law 
${\cal S}_\rho\propto (N_{\rm f})^{-\nu}$ with an exponent $\nu\approx 0.6$. 
The value of the exponent is essentially independent of the parameters $u$ and $W/t$, but depends 
on the confinement: for example, we find $\nu\approx 0.4$ for an open lattice with $V_2=0$. At high density ${\cal S}_\rho$ appears instead to be essentially unaffected by disorder.

The low-density behavior of ${\cal S}_\rho$ is reminiscent of what has been found in 
Refs.~\onlinecite{compressibility_anomaly} and~\onlinecite{droplet_state} for a 
$2D$ electron liquid. Following Ref.~\onlinecite{droplet_state} we explain the origin of the anomaly using the concept of density percolation. As $N_{\rm f}$ decreases the high-density regions tend to become disconnected, since the atoms tend to occupy just the deepest valleys in the disorder landscape. At given $u$ and low $N_{\rm f}$, the system thus stiffens as the disorder grows (see the bottom panel in Fig.~\ref{fig:four}). For a given realization of disorder, the number ${\cal N}_0$ of essentially empty sites increases with $W/t$, as it is shown in the inset.
 
There is, however, an important conceptual difference between the present Fermi gas and the $2D$ electron liquid. 
In the latter the density is also an inverse measure of the 
coupling strength: the stiffness anomaly at 
low density occurs in the strongly correlated regime. 
In the present case the atom number $N_{\rm f}$ and 
the interaction parameter $U$ are instead independent parameters.
The anomaly that we observe occurs also in the noninteracting limit 
(see the bottom panel of Fig.~\ref{fig:four}), demonstrating the crucial role of the quenching of percolation in originating the anomaly. The interatomic repulsions enhance the stiffness at low density in the disordered case just as they do in the clean case, in accord with the intuitive expectation that a repulsive system is less compressible.

In summary, we have shown how disorder affects the quantum phases 
of interacting Fermi gases moving under harmonic confinement 
in $1D$ lattices. In particular we have seen that Mott 
insulating regions are quite stable against uniformly distributed 
uncorrelated disorder and that the disorder induces an anomalous increase of the stiffness at low density from quenching of percolation.

This work was partially supported by an Advance Research Initiative of S.N.S. and by TUBITAK and TUBA.

\end{document}